\documentclass[pss]{wiley2sp} 

\usepackage[english]{babel}
\usepackage{amssymb,amsmath}
\usepackage{color}
\usepackage{epsfig}
\usepackage{graphicx} 
\usepackage{ulem}
\usepackage{cite}

\begin{document}

\title{Influencing the conductance in biphenyl-like molecular junctions with THz radiation}

\titlerunning{Short title }

\author{%
  Matthias Hinreiner\textsuperscript{\Ast,\textsf{\bfseries 1,2}},
  Dmitry A. Ryndyk\textsuperscript{\textsf{\bfseries 1,3}},
  Denis Usvyat\textsuperscript{\textsf{\bfseries 2}}, 
  Thomas Merz\textsuperscript{\textsf{\bfseries 2}},
  Martin Sch\"utz\textsuperscript{\textsf{\bfseries 2}},
  Klaus Richter\textsuperscript{\textsf{\bfseries 1}}}

\authorrunning{First author et al.}

\mail{e-mail
  \textsf{Matthias.Hinreiner@physik.uni-regensburg.de}, Phone: +49 941 943 3279, Fax: +49 941 943 4719}

\institute{%
  \textsuperscript{1}\, Institute for Theoretical Physics, University of Regensburg, Regensburg, Germany \\
  \textsuperscript{2}\, Institute for Physical and Theoretical Chemistry, University of Regensburg, Regensburg, Germany \\
  \textsuperscript{3}\, Institute for Materials Science and Max Bergmann Center of Biomaterials, Dresden University of Technology, Dresden, Germany}

\received{XXXX, revised XXXX, accepted XXXX} 
\published{XXXX} 


\abstract{%
%
%
%
\abstcol{
We investigate the torsional vibrations in biphenyl-like molecular junctions and transport properties in the presence of an external THz field. Ab-initio calculations including external electric fields show that the torsional angle $\phi $ of a thiolated biphenyl junction exhibits virtually no response. However, if  functional groups  are added to the molecule, creating a dipole moment in each of the rings, an external field becomes more effective for changing $\phi$. A model based on the $\cos^2\!\phi$ dependence of the current is proposed for the biphenyl-like molecular junctions in presence of an external THz field including 2,2'-bipyridine, 3,3'-bipyridine and 2,2',4,4'-tetramethyl-3,3'-bipyridine. The current through these molecules is shown to change if the THz frequency gets in resonance to the torsional vibration mode. 
}{}
}

%
%
\titlefigure{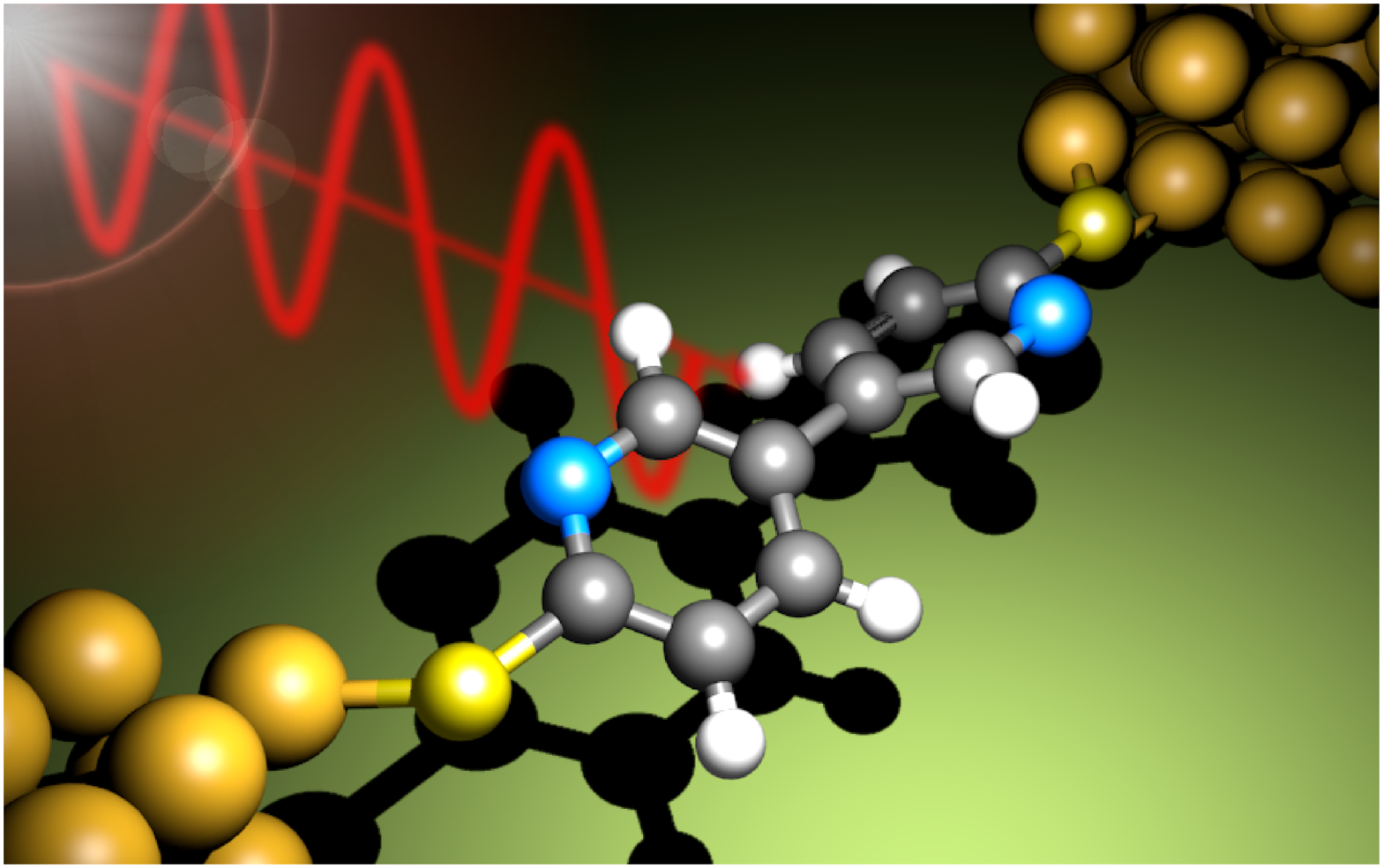}
\titlefigurecaption{%
  Dithiolated bipyridine between gold leads in the presence of external THz radiation.}

\maketitle   

\section{Introduction}

Molecular electronics is a field of research at the intersection of physics, chemistry, biology and engineering \cite{cuevas}. Since the pioneer work by Aviram and Ratner in 1974 \cite{Aviram1974277} great progress has been achieved in this field -- both theoretically and experimentally \cite{cuevas,richter}.  The evolution of molecular electronics is constantly stimulated by the search for new and better molecular devices that can reproduce the characteristics of electronic components like rectifiers and diodes \cite{doi:10.1021nl0352088,doi:10.1021/jp209547a,PhysRevLett.70.218,Elbing21062005}, switches and transistors \cite{natmater:blum,Liljeroth31082007,ADMA:ADMA201004291}, etc., but are smaller, faster and cheaper than their Si-based counterparts \cite{cuevas}.

Systems that constantly attract interest in this context are biphenyl (BPh) and biphenyl-like molecules \cite{Almenningen198559,arulmozhiraja:10589,Goeller2000399,kondo:064701,NatureVenkataraman,doi:10.1021/nl903084b,PhysRevB.77.155312,PhysRevB.53.R7626,doi:10.1021/j100154a030,PhysRevB.82.085402,doi:10.1021/ja107340t,doi:10.1021/jp065120t,epub1745}, since they represent relatively simple molecular junctions that, however, through their torsional modes allow for investigating interplay between charge transport and molecular vibrations.  In this work the focus is on biphenyl-like molecular junctions with metal electrodes and their behavior in the presence of an external THz radiation. This is of interest from several points of view. Every investigation of molecular junctions and their response to external perturbations could help for a better understanding of molecular transport. The investigation of THz radiation's influence on molecular junctions could provide insight to molecular vibrations and their effect on conductance or conformational switching.
 
Another interesting point about biphenyl-like junctions is that they may allow one to build-up a THz-sensitive detector. In the last years there has been considerable progress in finding molecular detectors for electro-magnetic radiation in the visible range \cite{PhysRevLett.91.207402,0957-4484-16-6-012,PhysRevLett.93.248302,Choi2001241}.  In this case the electronic transfer is directly coupled to phonons, and a fully quantum mechanical treatment with electron-phonon terms in the Hamiltonian is required \cite{cuevas}. However, in the THz case vibrations correspond to the timescale of picoseconds, which is much slower than the femtosecond electronic transfer processes\cite{Galperin:1,0953-8984-19-10-103201}.  Therefore, the vibrations in the THz case can be treated as classical oscillations of the molecular geometry. This allows for decoupling of the electronic transfer within the molecular junctions from the relatively slow geometry changes due to the THz vibrations.

The paper is organized as follows. In Sec. 2 we consider a model for biphenyl-like molecular junctions in the presence of THz radiation and discuss the relevant parameters. The results of our calculations for biphenyl and two other biphenyl-like molecules are given in Sec. 3, which ends with a discussion on increasing the sensitivity on THz radiation by using appropriate molecules. In Sec. 4 the final conclusions are given.

\section{Torsional vibrations in THz field}
\label{sec:THz}

\begin{figure}[t]
  \centering
  \includegraphics[width=0.4\textwidth,keepaspectratio]{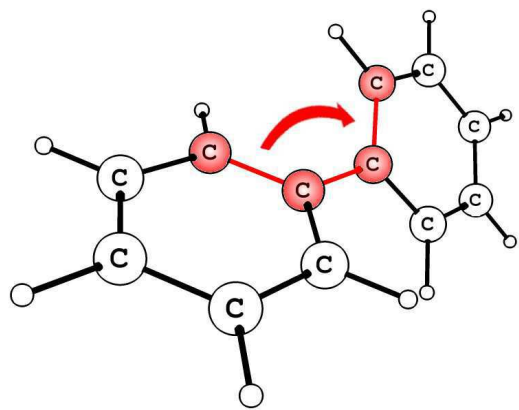}
  \caption{Biphenyl molecule (BPh) and the definition of the torsional angle $\phi$ as the dihedral angle specified by the four red C-atoms.}
  \label{biphenyl_definition}
\end{figure}

In this section we show how a biphenyl-like molecular junction with gold leads responds to an external THz radiation. As it is generally known, and is also explicitly shown in sections \ref{basics} and \ref{sec:Cos2law} below, the conductance of a BPh junction substantially depends on the torsional angle between the two phenyl rings (see fig. \ref{biphenyl_definition}). Therefore, by controlling the torsional angle, one can influence the conductance of a junction.

One known possibility to manipulate the torsional angle of BPh is to induce a molecular charge. According to B3LYP calculations by Arulmozhiraja et al.\ \cite{arulmozhiraja:10589}\ the torsional angle of neutral BPh is $40.1^\circ$ whereas in the negatively charged ion it becomes $0^\circ$ and in the positively charged one $18.9^\circ$.  Thus, a change of the charge state of the molecule, e.g.\ by applying a gate voltage, leads to a sharp change of the conductance.

In this paper we consider an alternative mechanism. A THz electric field can cause resonant excitation of the molecular torsional vibrational mode $\Omega_0$ if their frequencies are close, which leads to a change in the conductance through the junction. \cite{Lehmann2003282}

\subsection{Basic properties of biphenyl-like junctions}
\label{basics}

Before we consider THz excitations of biphenyl-like junctions we introduce the ``$\cos^2(\phi)$ law" of its conductance.  In biphenyl-like molecular junctions the current is mainly carried via the $\pi$-electrons of the aromatic rings \cite{PhysRevB.53.R7626}.  The conductance through the BPh junction substantially depends on the torsional angle $\phi$, since in planar BPh the $\pi$-orbitals of both rings are strongly coupled, whereas for $\phi=90^\circ$ they are almost decoupled from each other.  In the latter case the charge transfer from one ring to the other is essentially suppressed compared to the former one resulting in a much lower conductance.

Many theoretical, computational as well as experimental studies have investigated the dependence of the zero-voltage conductance $G|_{V=0}$ through BPh and biphenyl-like junctions on its torsional angle and found the approximation \cite{PhysRevB.82.085402,PhysRevB.77.155312,doi:10.1021/nl903084b,doi:10.1021/ja107340t}
\begin{equation}
  G|_{V=0} \propto \cos^2 ( \phi)
  \label{cos2eq}
\end{equation}
to be reasonable accurate. This relation can be understood in two ways. First, the current through \textsl{different} biphenyl-like molecules depends on the $\cos^2(\phi_0)$ of the torsional angle $\phi_0$ of the fully optimized geometry. For example\cite{NatureVenkataraman} and \cite{doi:10.1021/nl903084b} investigated the relation in this context.  Second, and more important for this work, the conductance of \textsl{one specific} molecular junction as a function of the torsional angle $\phi$ behaves according to $\cos^2(\phi)$. Kondo et.\ al., for example, investigated the conductance dependence of BPh on its torsional angle this way \cite{kondo:064701}.

As second approximation entering our model we assume, that the dependence of the potential energy of biphenyl-like junctions can be well described by a harmonic potential around the torsional angle of the optimized molecule $\phi_0$. For small deviations from the minimum $\phi_0$ this is certainly fulfilled. However we will assume it to hold for all relevant torsional angles, which will be needed to derive the model equations below.  In section \ref{subsec:Ground state geometry and energy} and \ref{sec:Cos2law} both assumptions will be verified.

\subsection{The model}

The assumption of a harmonic potential with the energy minimum at $\phi_0$ allows one to describe the dynamics of the junction's torsional angle by a classic harmonic oscillator.  At this point the validity of Born-Oppenheimer approximation is assumed separating electronic motion and the movement of nuclei. The classical treatment of nuclear dynamics can be legitimated by our aim to focus on low-frequency oscillations only.  To achieve an external angle control through the excitation of torsional vibrations an external time-dependent driving force is considered.  This is achieved by means of a harmonic electric radiation field $F\propto f \cos(\omega t)$, leading to the equation of motion
\begin{equation}
 \ddot \varphi + \Gamma \dot \varphi + \Omega_0^2 \varphi = f \cos(\omega t).
 \label{EoM with drive} 
\end{equation}
Here $\varphi$ denotes the displacement from the minimum $\phi_0$ (i.e. $\varphi=\phi-\phi_0$), $\Gamma$ is a phenomenological damping parameter of the harmonic torsional vibration with frequency $\Omega_0$.  The parameters of the driving force are $\omega$, the angular frequency of the external electric radiation, and $f$, a response parameter of the molecule with respect to the electric field of a given strength.  The direction of the field is assumed to be perpendicular to the axis of the inter-ring bond of the molecule,
since otherwise it would result in an undesirable modulation voltage on top of the constant bias voltage.

Equation (\ref{EoM with drive}) can be solved analytically yielding
\begin{align}
  \label{solution of EoM}
 \varphi(t) & = \Phi_h \mbox{e}^{\left(-\frac{\Gamma}{2} t \right)} \cos\left(\omega t + \Psi_h \right)+ \Phi \cos(\omega t + \Psi), 
\end{align}
with
\begin{align}  
\label{equation for Psi}
 \Psi & = \arctan \left\{ \frac{\Gamma \omega}{\omega^2 - \Omega^2} \right\}, \\
  \label{equation for Phi}
 \Phi & = \frac{f}{\sqrt{\Gamma^2 \omega^2 + {\left( \omega^2 - {\Omega_0}^2 \right)}^2}},  
\end{align}
and $\Phi_h$ and $\Psi_h$ being free parameters depending on the initial or boundary conditions.

For long times or large damping rates, i.e. $\Gamma t>>1$, the solution of equation (\ref{EoM with drive}) is approximately
\begin{equation}
 \varphi(t)  =  \Phi \cos(\omega t + \Psi) 
  \label{large t solution of EoM}
\end{equation}
as the first term in equation (\ref{solution of EoM}) is exponentially suppressed.

In view of equation (\ref{cos2eq}) the conductance of a biphenyl-like junction depends on its torsional angle
\begin{align}
 G_{\omega}(t) \propto \cos^2\left(\phi_0+ \Phi \cos(\omega t + \Psi) \right).  
 \label{G_forumla_1}
\end{align}

In the linear response regime of small bias voltage the current through the molecular junction holds the same dependency.  If the frequency $\omega$ of the external driving field is fast compared to typical switching times of electric engineering, i.e.\ MHz or GHz, the effective current averaged over the time period $T=\frac{2\pi}{\omega}$,
\begin{equation}
 {\left< I\right>_T }^{\omega}\propto \frac{\omega}{2\pi} \int_0^{\frac{2\pi}{\omega}}\! \mathrm{d}t \, \cos^2 \! \left( \phi_0+\Phi \cos \left( \omega \, t \right) \right),
    \label{harmonic current approx} 
\end{equation}
is of more relevance than the time-resolved current $I(t)$.
The integral in equation (\ref{harmonic current approx}) can be solved analytically using the integral representation of the zeroth-order Bessel function \cite{gradshteyn}:
\begin{align}
 \int_0^{2\pi}  \cos(2\Phi\sin\tau)\,\mathrm{d}\tau &= 2\pi J_0(2\Phi),
\end{align} 
yielding
\begin{equation}
 {\left< I\right>_T }^{\omega}\propto \frac{1}{2}  \,+\,\frac{\cos (2\phi_0)}{2} \, \mbox{\large \it J}_0 (2\Phi).
  \label{solution effective current}
\end{equation}

For a more general $\left|\cos(\phi)\right|^{\alpha}$ dependence of the conductance, the corresponding integration (\ref{harmonic current approx}) has usually to be performed numerically.

\subsection{Parameters of the model}
\label{sec:parameters}

Here, we describe the methods to obtain the model parameters, entering equations (\ref{equation for Phi}) and (\ref{solution effective current}).

The {\bf torsional angle $\boldsymbol{\phi_0}$}, corresponding to the fully optimized molecule geometry was obtained by DFT electronic structure calculation (see below), and corresponds to the minimum of the potential energy surface.

The {\bf frequency of the torsional vibration $\boldsymbol{\Omega_0}$} within harmonic approximation was also calculated at DFT level.

The {\bf response parameter $\boldsymbol f$} describes how strong the molecule reacts to the presence of the electric radiation of a given strength.  Since this parameter is assumed to be frequency independent, it is obtained by analyzing the molecule's response to static fields.  For a time-independent external force the right hand side of equation (\ref{EoM with drive}) is constant and the equation of motion reduces to
\begin{equation}
 {\Omega_0}^2\varphi=f,
\label{formula for f}
\end{equation}
where $\varphi$ is the angle change due to the static external field of a given strength. The corresponding finite field calculations were also performed at the DFT level.

Evaluation of the {\bf damping of the torsional oscillation $\boldsymbol \Gamma$} is more involved.  To this end we use a molecular dynamics simulation to investigate the behavior of the molecular junction without external driving but with a non-equilibrium initial value, i.e. $\varphi(t=0)\ne0$, and analyze how the torsional angle decays in time.  The damping parameter $\Gamma$ can be then obtained by a fit of the function $\phi(t)=\Phi_h \cos(\Omega_0^\prime t + \Psi_h)\exp(-\frac{\Gamma}{2}t)+\phi_0^\prime$ to the resulting curve obtained in the dynamics calculation, i.e.  the solution of the damped harmonic oscillator without external driving. This involves five parameters: $\Phi_h$, $\Psi_h$, $\Gamma$, $\Omega_0^\prime$ and $\phi_0^\prime$. The last two should ideally correspond to $\Omega_0$ and $\phi_0$ obtained by direct calculations (vide supra).

For making calculations affordable several approximations were applied.  First, it is assumed that $\Gamma$ is independent of the oscillation's amplitude $\Phi$, as is already implied in equation (\ref{EoM with drive}). Second, the torsional mode is assumed to be not coupled to the leads directly, but only to other vibrational modes of the molecule, which in turn are well coupled to the leads.  Therefore the energy of the torsional vibration excited by external radiation dissipates only through other vibrational modes.  Within this approximation all modes except the torsional one are almost in thermal equilibrium because they can effectively transfer any heat to the electrodes.  This allows us to reduce each of the gold electrodes to only one single atom fixed in space and assume that the damping parameter found this way is approximately of the same order of magnitude as the one for the real junction.

{To model this process we performed a series of calculations.  An NVT (canonical ensemble) calculation is primarily done until thermal equilibrium is achieved. The starting point is the optimized molecule except that the torsional angle is highly displaced. Thus the most dominant motion in thermal equilibrium is the torsional oscillation. The temperature used for the NVT calculation is set to a value, for which the corresponding torsional amplitude is approximately $5^\circ$.  Then from this simulation the geometry of the molecule was taken at a point in time when the torsional angle is at a turning point.  That geometry was used for the actual NVE (microcanonical ensemble) dynamics calculation, employed for the evaluation of $\Gamma$ via the fit.}

Finally, the {\bf amplitude $\boldsymbol \Phi$} of the torsional angle oscillation is calculated from equation (\ref{equation for Phi}).

Further,we define two more parameters:

The {\bf amplitude in resonance, $\boldsymbol {\Phi_{\mbox{\tiny{max}}}} $,} is the maximal possible amplitude for a specific junction combined with radiation of a given strength. According to equation (\ref{equation for Phi}) it is obtained as
  \begin{equation}
   \Phi_{\mbox{\tiny{max}}}  = \frac{f}{\Gamma \cdot \Omega_0}.
    \label{Phi max}
  \end{equation}
  
  The {\bf relative current change in resonance, $\boldsymbol{\Delta_{\mbox{\tiny{rel}}}^{\omega}I}$,} is defined as
  \begin{equation}
   \Delta_{\mbox{\tiny{rel}}}^{\omega}I := \frac{{\left< I\right>_T }^{\omega}-{\left< I\right>_T }^{0}}{{\left< I\right>_T }^{0}},
  \label{basic relative current change equation}
  \end{equation}
where ${\left< I\right>_T}^{\omega}$ is the time-averaged current through the junction in presence of an external field with frequency $\omega$ (eqs. (\ref{harmonic current approx}) and (\ref{solution effective current})).

\section{Results and discussion}
\label{results}

\subsection{Molecules}

In this subsection we describe shortly the junction molecules, used in the calculations.
Biphenyl (BPh) is included to serve as reference despite its ineffectiveness for THz-sensitive molecular junctions (as we will see in Sec. \ref{sec:Influence of external electric fields}). The second molecule, 3,3'-bipyridine (BPy3), is chosen for the similarity of most of its parameters to those of BPh's apart from the sensitivity to electric fields. Molecules like 2,2'-bipyridine (BPy2) are less ``biphenyl-like", but can still be used within our formalism.

\begin{figure}[t]
  \centering
  \includegraphics[width=0.49\textwidth,keepaspectratio]{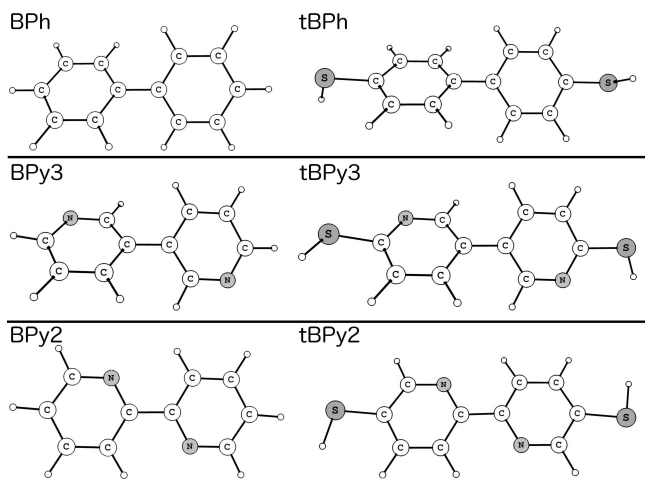} \\[0.2cm]
  \caption{Biphenyl (BPh), 3,3'-bipyridine (BPy3), 2,2'-bipyridine (BPy2) and thiolated counterparts.}
  \label{biphenyl_formula}
\end{figure}

Figure \ref{biphenyl_formula} shows the structural formula of all three molecules and additionally thiolated ones as sulfur will be used as bridging atom to connect the molecules to the gold leads in our test calculations.

\subsection{Computational parameters}

The ab-initio calculations were performed, unless specified differently, with TURBOMOLE \cite{TURBOMOLE} or the Firefly QC package \cite{firefly}{, which is partially based on the GAMESS (US) \cite{pcgamess} source code. } Test calculations on biphenyl-like molecules showed a considerable change of the results by increasing the basis set from split-valance to triple zeta, whereas a quadruple-zeta basis does provide almost no further improvement.  Several functionals (namely BLYP, PBE~96 and B3LYP) have been tested, however, the influence of the function on the relevant results is found to be tiny.

\subsection{Ground state geometry and energy}
\label{subsec:Ground state geometry and energy}

The geometry of BPh has already been investigated by other groups, both theoretically and experimentally \cite{Almenningen198559, arulmozhiraja:10589, doi:10.1021/j100154a030}.  Although there are differences in the exact values of the parameters at different levels of calculations, they all show that the torsional angle between the almost planar rings is roughly about $35^\circ - 45^\circ$ on average slightly below the values $44.4\pm1.2^\circ$ \cite{Almenningen198559} and $40^\circ-43^\circ$ \cite{19591340} of experiment.
      
      \begin{figure}[b]
          \centering
          \includegraphics[width=0.5\textwidth, keepaspectratio]{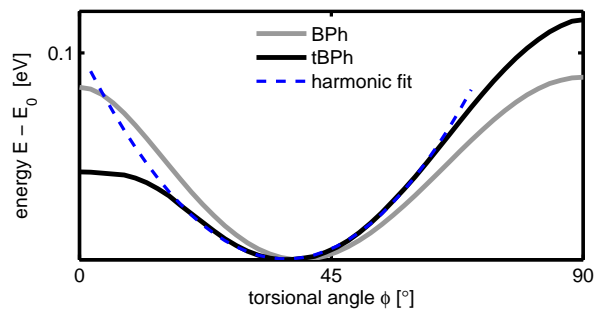}
          \caption{Ground state energy of BPh and tBPh depending on the torsional angle.}
          \label{BPh-SH angle energy}
      \end{figure}

The gray curve of figure \ref{BPh-SH angle energy} shows BPh's rotational energy curve for torsional angles between $0^\circ$ and $90^\circ$ calculated in $2^\circ$-steps.  At a torsional angle of $\phi_0=40^\circ$ the energy exhibits its minimum as expected. The general curve as well as the values of the rotational barrier, $\Delta E^0$ and $\Delta E^{90}$, are in good agreement with previous work \cite{arulmozhiraja:10589, Goeller2000399}.
      
In order to estimate the influence of the leads on the BPh molecule we performed a potential energy scan also for tBPh.  The results are shown by the black line in figure \ref {BPh-SH angle energy}.  Addition of sulfur slightly modifies the optimized value of the torsional angle by $4^\circ$ yielding $\phi_0=36^\circ$.  Despite the substantial changes in the heights of the rotational barriers, the important region around $\phi_0$ is less influenced. A previous study by Pauly et.\ al.\ \cite{PhysRevB.77.155312} also demonstrates that effect on the rotational barriers by adding sulfur. In that work, however, the geometry parameters for torsional angles away from the minimum were not optimized, which resulted in breaking mirror symmetry with respect to $\phi=90^\circ$.
      
A further mimicking of the leads (namely by adding the two Au atoms closest to the molecule) have only a negligible impact on the region close to $\phi_0$. Hence we reduce the leads to only the bridging sulfur atom in the calculation of most molecular properties.
      
The harmonic fit of the tBPh curve (blue dashed line in figure \ref{BPh-SH angle energy}) around the minimum $\phi_0$ shows that the curve can be well described by a harmonic potential for a region of approximately $\phi\pm20^\circ$.
   
\begin{figure}[htb]
  \centering
  \includegraphics[width=0.5\textwidth, keepaspectratio]{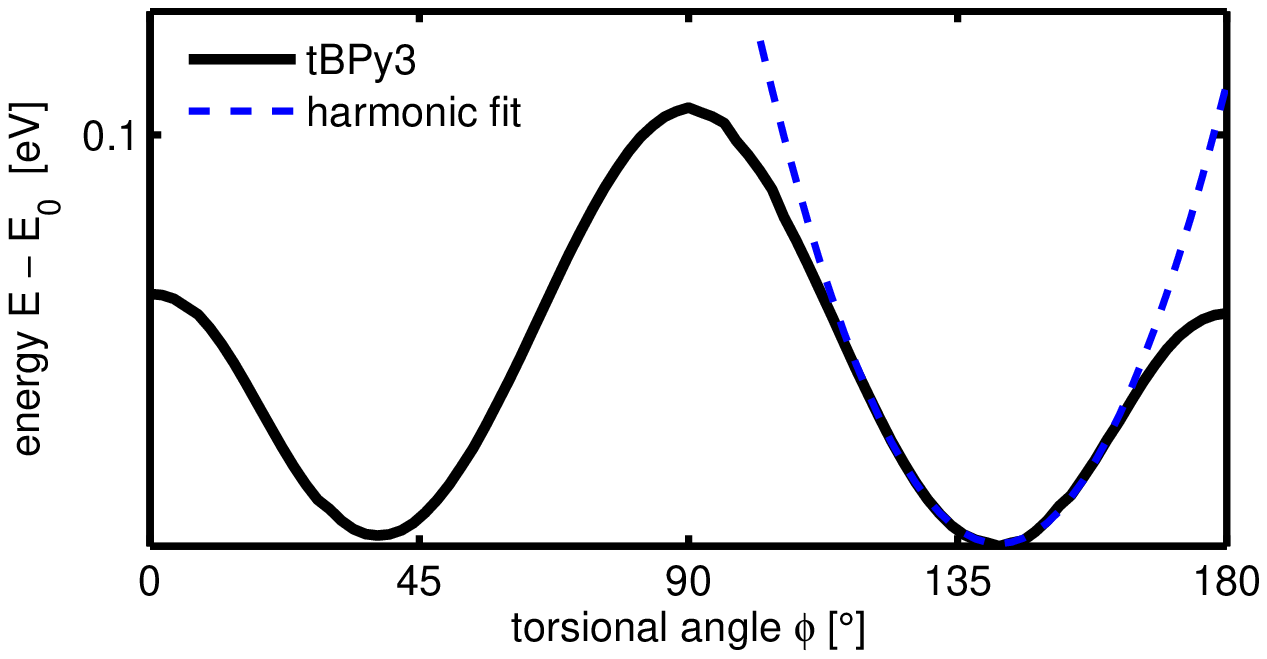}
  \includegraphics[width=0.5\textwidth, keepaspectratio]{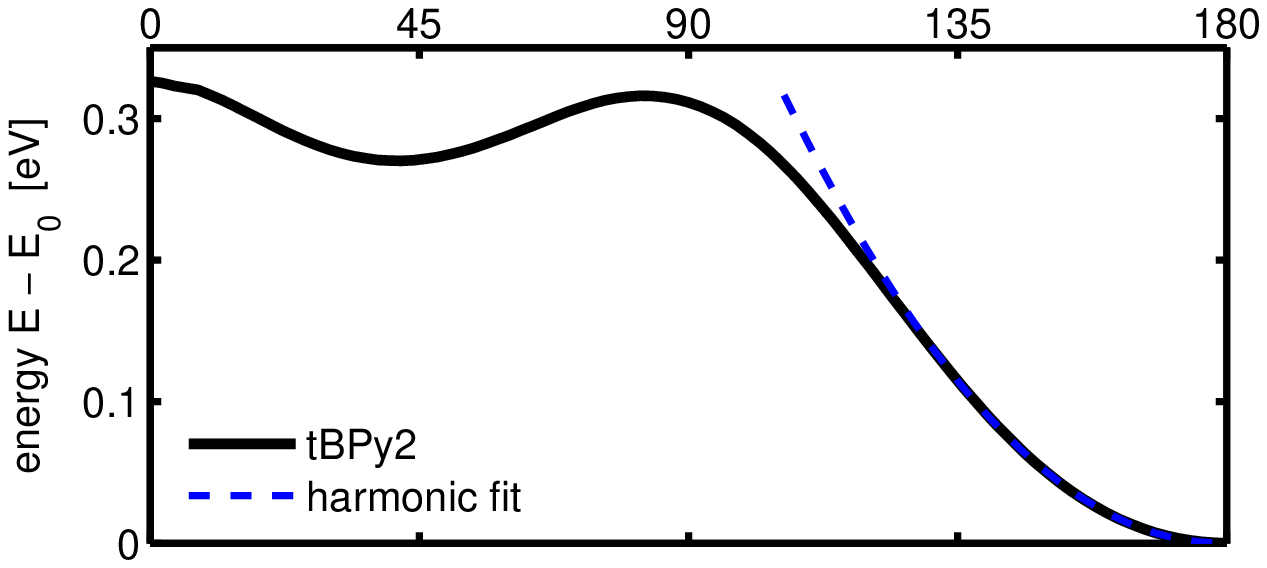}
  \caption{Energy surface depending on torsional angle for tBPy3 and tBPy2.}
  \label{BPy3 angle energy}
 \end{figure}

BPy3 and tBPy3 are slightly closer to a perpendicular structure than their biphenyl counterparts (by not more than $2^\circ$).  Also the energy dependence on the torsional angle of tBPy3 is close to that of tBPh, yet tBPy3's tendency towards the perpendicular structure can be seen at the torsional barrier heights, too: The energy difference between the planar tBPy3 and the optimized one is 50\% higher than that of tBPh, whereas $\Delta E_{90}(\mbox{tBPy3})$ is 0.01~eV lower than $\Delta E_{90}(\mbox{tBPh})$.  Figure \ref{BPy3 angle energy} shows the calculated potential curve for tBPy3.  Due to tBPy3's (and also tBPy2's) reduced symmetry compared to tBPh the torsional angle ranges from $0^\circ$ to $180^\circ$. A torsional angle of $\phi=0^\circ$ correspond to the case with both nitrogen atoms on the same side (cis configuration), whereas for $\phi=180^\circ$ they oppose each other (trans configuration). The global minimum at $142^\circ$ is 2.6\ meV lower than the local one at $\phi=38^\circ$.  This difference can be explained by mutual repulsion between the two nitrogen atoms.  It is quite small compared to the rotational barrier heights ($\Delta E^{0,\, 90} > 60~\mbox{meV}$) and therefore BPy3 with geometries corresponding to both minima should be observed in experiments.

For BPy2 and tBPy2 the interaction between the N-atoms with the HC-groups of the adjacent rings (see figure \ref{biphenyl_formula}) lowers considerably the energy for the planar structure \cite{Goeller2000399}. This effect dominates the optimized torsional angle, which in this case is $\phi_0=180^\circ$. The corresponding potential energy curve is shown in figure \ref{BPy3 angle energy}.  Yet, for angles below $90^\circ$, for which that interaction is reduced, the curve shows a behavior similar to tBPh and tBPy3.

The blue dashed lines represent the approximated harmonic potentials for tBPy3 and tBPy2, which recover the potential energy curves well in the range of at least $\phi_0\pm20^\circ$ for tBPy3 and of about $\phi_0\pm55^\circ$ for tBPy2.

\subsection{Influence of external electric fields}
\label{sec:Influence of external electric fields}

For the calculation of the $f$ parameter according to section \ref{sec:parameters} the torsional angles of BPh, BPy3 and BPy2 are calculated for several field strengths and directions perpendicular to the inter-ring bond of the external static field.

The torsional angle of BPh, whose rings do not have effective permanent dipole moments perpendicular to the inter-ring bond, is hardly affected by fields up to 1\ V/nm. At the same time, $\phi_0$ of BPy3 and BPy2 shifts nearly linearly with increasing field strength. For 0.1\ V/nm the optimized torsional angles change in BPy3 and BPy2 up to $0.30^\circ$ and $0.48^\circ$, respectively, depending on the orientation of the molecule.  We have chosen to take the $f$ value obtained by a field of 0.1\ V/nm, because, on the one hand, the field should be strong to maximize the effect, but, on the other hand, should not go far beyond potential experimental possibilities. 

\subsection{Torsional modes}
\label{sec:tors}

Next, we investigate the vibrational modes of BPh, BPy3 and BPy2 in the junction.  As already pointed out in section \ref{sec:parameters}, the intra-lead vibrations, which are costly to calculate, can be omitted from consideration. Therefore, we reduce them to only one Au-atom with its mass increased to $m_{Au}= 10^6\, \mbox{u}$.
 
\begin{figure}[t]
  \centering
  \includegraphics[width=0.35\textwidth, keepaspectratio]{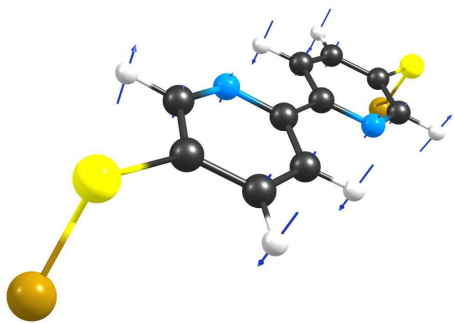}
  \caption{Torsional vibration of BPy2.}
  \label{BPy_torsional vibration}
\end{figure}

The calculated harmonic vibrational frequencies range from about $10$ cm$^{-1}$ for the collective oscillation of the whole molecule relatively to the leads (represented by the heavy gold atoms) to several thousand cm$^{-1}$ corresponding to fast oscillations of the light H-atoms.  Of major importance is the frequency $\Omega_0=2\pi\cdot\nu$ of the torsional vibration, which is sketched in figure \ref{BPy_torsional vibration} for BPy3 (for BPh and BPy2 it is analogous).  The calculated values for this frequency are $\nu^{\mbox{\tiny{BPh}}}= 2.2\mbox{ THz}$ for BPh, $\nu^{\mbox{\tiny{BPy3}}}=2.3$\ THz for BPy3 and $\nu^{\mbox{\tiny{BPy2}}}=1.7$\ THz for BPy2.
    
\subsection{Verification of $\cos^2 \phi$ law}
\label{sec:Cos2law}

In the section we demonstrate the ``$\cos^2(\phi)$-law'' (see section \ref{basics}) for BPh and calculate the actual conductance dependency of BPy3 and BPy2 on the torsional angle.

      \begin{figure}[htb]
      \centering
      \includegraphics[width=0.5\textwidth, keepaspectratio]{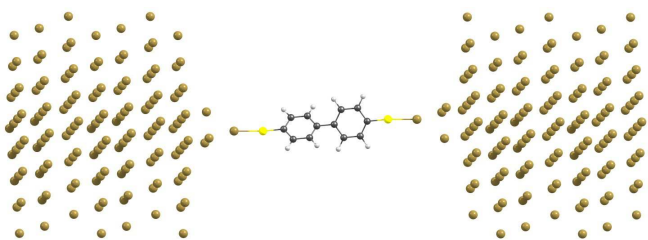}
      \caption{System used for conductance calculation of BPh junctions. For BPy3 and BPy2 correspondingly.}
      \label{BPh_conduct_system}
    \end{figure}

For the transport calculation the ``DFTB$^+$ NEGF'' code \cite{doi:10.1021/jp070186p, 1367-2630-10-6-065022} is used.  It is based on the Density Functional Tight Binding method combined with non-equilibrium Greens function techniques.

The considered geometry is depicted in figure \ref{BPh_conduct_system}.  It consists of two electrodes each build up with 120 Au-atoms in such way that it is compatible with periodic boundary conditions. The BPh, BPy3 or BPy2 molecule and the sulfur bridging atoms are in between and form together with 48 adjacent gold atoms he device region.

The current through the junction is calculated for several torsional angles in $10^\circ$-steps from $0^\circ$ to $180^\circ$. The remaining geometry parameters of the BPh molecule are mainly identical to those obtained by a DFT geometry optimization with a constrained torsional angle, however some minor adjustments had to be made to fit it properly in between the leads as they are fixed in space for all calculations.  The bias voltage is set to $U=0.01$~mV.  This value is somewhat arbitrary but it has to be chosen small to approximate $G|_{V=0}$.

    \begin{figure}[t]
      \centering
      \includegraphics[width=0.5\textwidth, keepaspectratio]{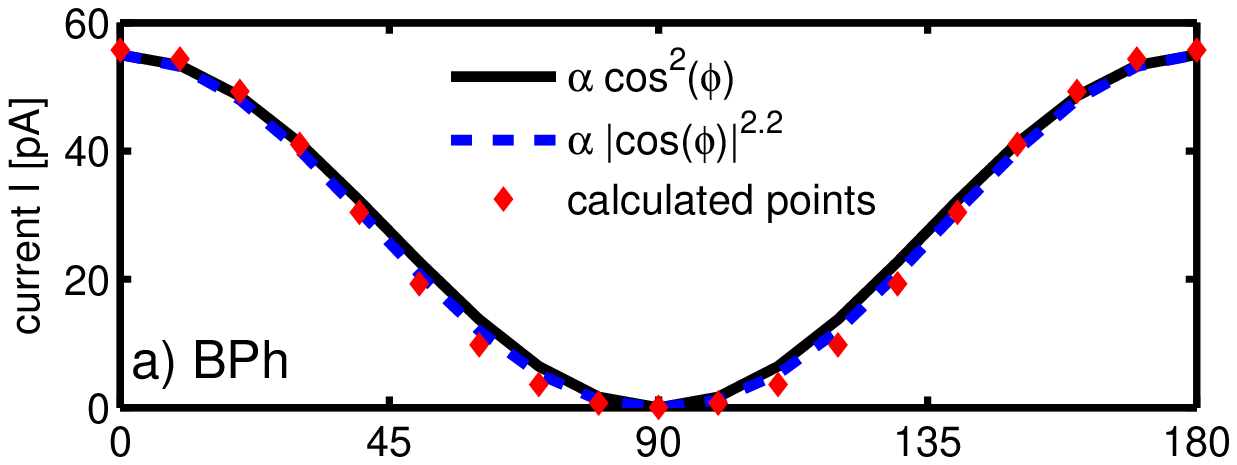}
      \includegraphics[width=0.5\textwidth, keepaspectratio]{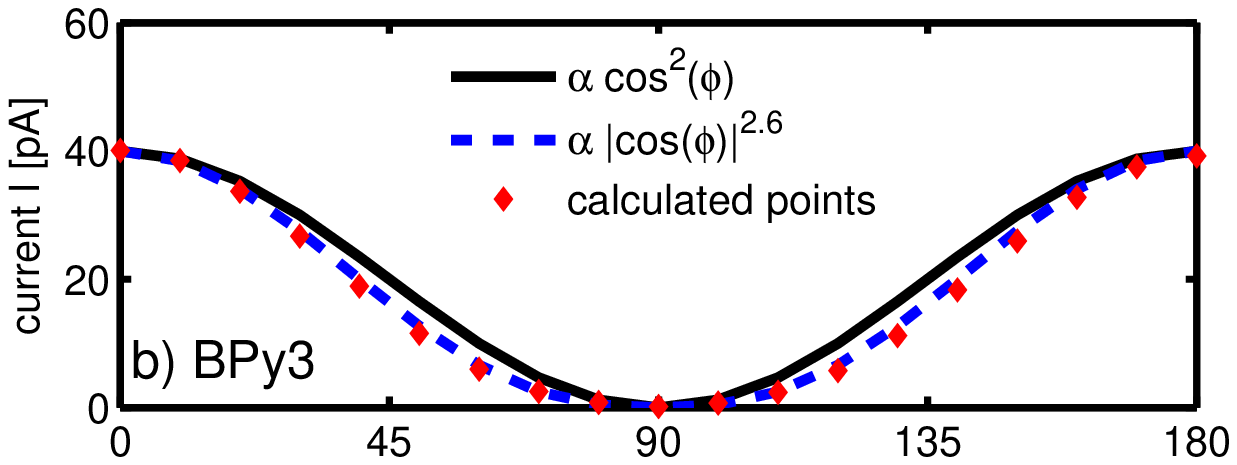}
      \includegraphics[width=0.5\textwidth, keepaspectratio]{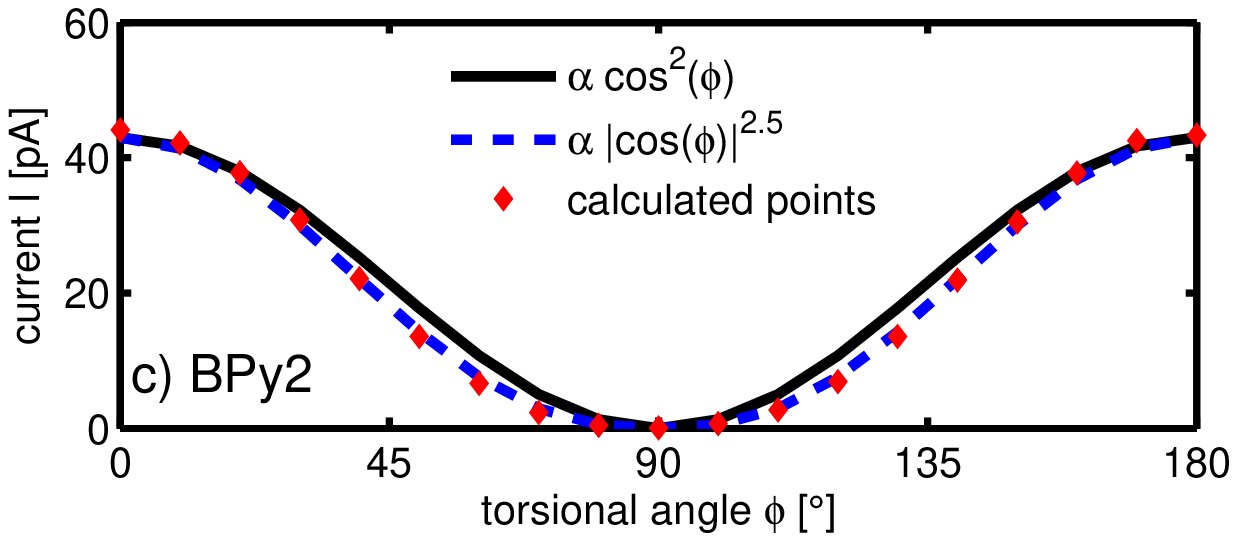}     
      \caption{Conductance through BPh (a), BPy3 (b) and BPy2 (c) junction depending on torsional angle $\phi$  for bias voltage $U=0.01$\ mV.}
      \label{BPh angle conductance}
    \end{figure}

The obtained results can be seen in figure \ref{BPh angle conductance}a. The plot shows the current $I$ instead of the conductance $G|_{V=0}\approx\frac{I}{U}$.  Red diamonds represent the results of individual calculations.

The values are in quite good agreement with the black solid $\cos^2(\phi)$-curve, however the values for $50^\circ$, $60^\circ$ and $70^\circ$ (as well as for $110^\circ$, $120^\circ$ and $130^\circ$ that are the same by symmetry) are somewhat below the $\cos^2(\phi)$ ones.  The lowest current value is $4.2\cdot10^{-4}$~pA obtained for $\phi=90^\circ$, the highest one is 56~pA for $\phi=0^\circ$ and $\phi=90^\circ$. Thus the ``max-min-ratio'' is $\frac{I_{\mbox{\tiny{max}}}}{I_{\mbox{\tiny{min}}}}>10^{5}$ and one can certainly assume $I(\phi=90^\circ)$ to be effectively zero in accordance with the $\cos^2(\phi)$ law.

The calculated data would suggest a little change of the exponent: $I\propto\left|\cos(\phi)\right|^{2.2}$ (blue dashed curve) fits almost perfectly the calculated values, but the maximal deviations to the $\cos^2(\phi)$ law are also less than 7\% of the highest current.

These results are in quite good agreement with those reported by Kondo et.\ al. \cite{kondo:064701}, who investigated the conductance through a BPh junction and its torsional dependency with a NEGF method based on DFT, and of Pauly et.\ al.\ \cite{PhysRevB.77.155312} who calculated the conductance in the Landauer-B\"uttiker formalism. Both of them report a $\cos^2(\phi)$-like dependency of the conductance, however for the max-min-ratio considerably lower values than ours: Pauly et.\ al. get a ratio of 227 and Kondo et.\ al. get 124.

The obtained data for BPy3 (and BPy2) are shown as the red diamonds figure \ref{BPh angle conductance}b (figure \ref{BPh angle conductance}c). The maximal currents of 40~pA (44~pA) is obtained as expected for $0^\circ$ and $180^\circ$; a torsional angle of $90^\circ$ leads to the minimal value of 0.18~pA (0.045~pA). The max-min-ratio is slightly higher than 200 (almost 1000).

Comparing these results to those of the BPh junction, one sees that their conductance in all systems behaves in general similarly. However, there are certain differences: For torsional angles away from $\phi=0^\circ$ the current through BPy3 and BPy2 is about one third weaker than through BPh.  The very weak current at $90^\circ$ for BPy3 and BPy2 is 2--3 orders of magnitudes stronger compared to BPh, resulting in a much lower max-min-ratio. The value for this ratio is actually of the order of that proposed for BPh in other studies \cite{kondo:064701,PhysRevB.77.155312}.

Another difference becomes obvious by comparing the calculated data with a $\cos^2(\phi)$-curve (solid black lines in figure \ref{BPh angle conductance}b) and c)). Although the shape of the current through BPy3 and BPy2 can be still described as ``$\cos^2(\phi)$-like'', the differences are yet larger than for BPh. An exponent of 2.6 and 2.5 (blue dashed curves in \ref{BPh angle conductance}b) and c)) would fit the data much better than the exponent 2 of the $\cos^2(\phi)$ law or 2.2 of the fit for BPh.

In the following sections we nevertheless assume that conductance through BPy3 and BPy2 obay the $\cos^2(\phi)$ law.

\subsection{Damping parameter}

 \begin{figure}[b]
  \centering
  \includegraphics[width=0.5\textwidth, keepaspectratio]{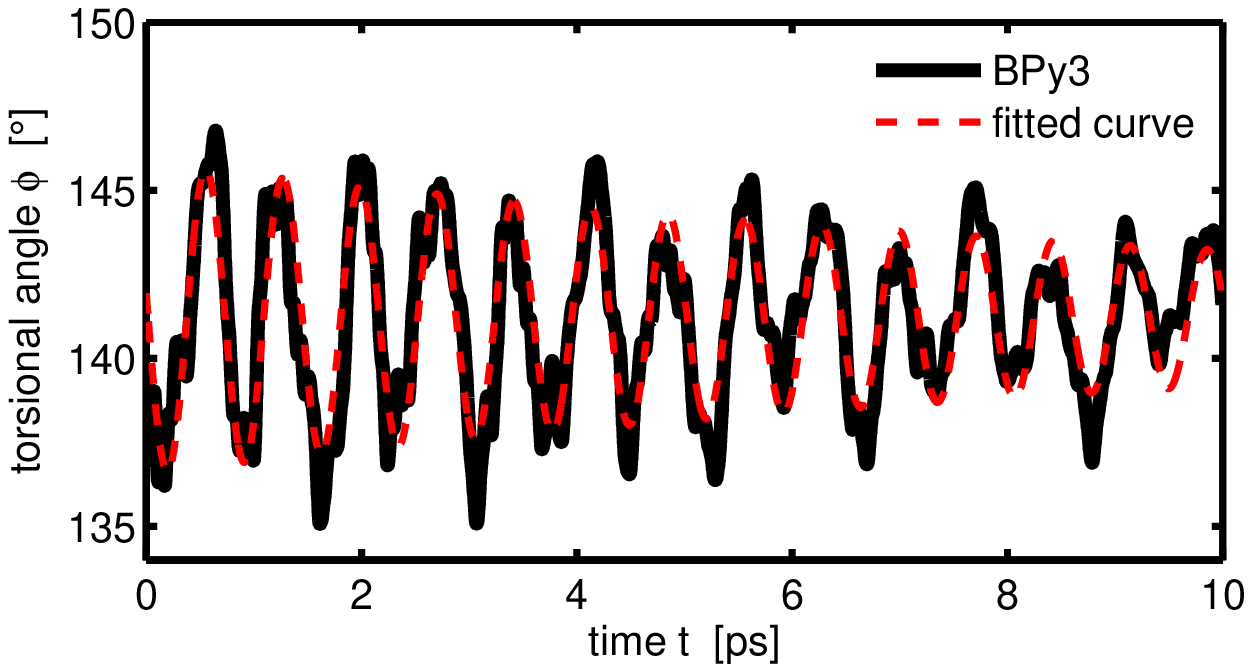}
  \includegraphics[width=0.5\textwidth, keepaspectratio]{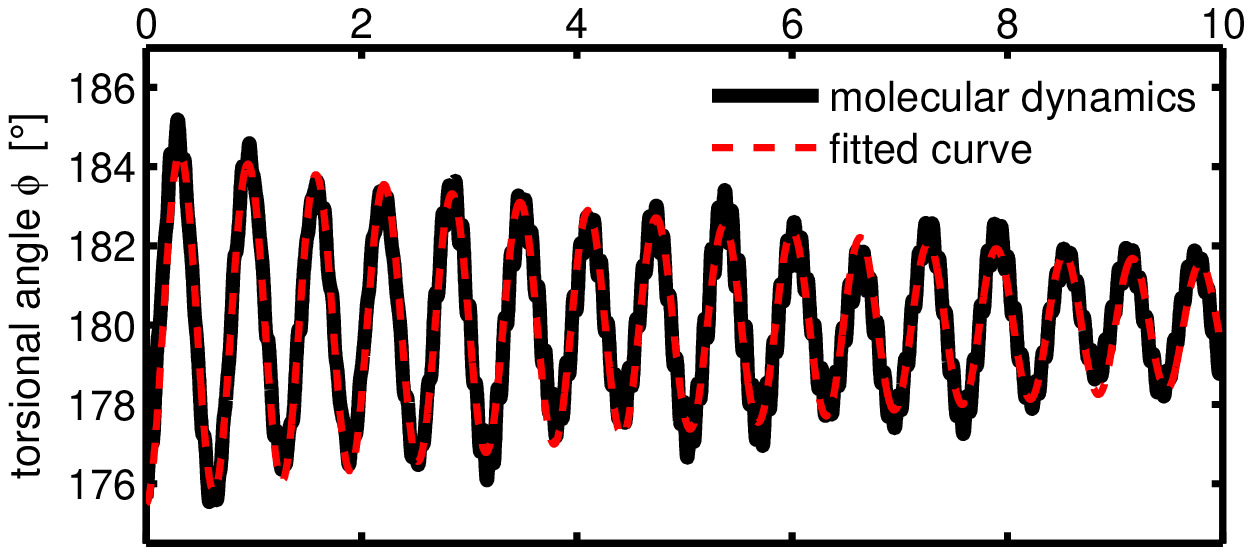}
  \caption{Results for the molecular dynamics calculation for BPy3 and BPy2.}
  \label{dynamcis}
 \end{figure}

The damping parameter $\Gamma$ is estimated for BPy3 and BPy2 by means of a molecular dynamics calculation according to section \ref{sec:parameters}. These results, together with the $\phi(t)=\Phi_h \cos(\Omega_0^\prime t + \Psi_h)\exp(-\frac{\Gamma}{2}t)+\phi_0^\prime$ fitting curve, are given in figure \ref{dynamcis}. The obtained values are: $\Gamma_{\mbox{\tiny{BPy2}}}=1.67\cdot10^{11}~\mbox{s}^{-1}$ and $\Gamma_{\mbox{\tiny{BPy2}}}=2.14\cdot10^{11}~\mbox{s}^{-1}$

The molecular dynamics calculation can additionally serve as a check for quality of the model equation and the torsional frequency $\Omega_0$. For BPy2 the fitted curve (i.e. the expect behavior as indicated by the model equation) reproduced almost perfectly the calculated values. Also the frequency of the oscillation, $\Omega_0^\prime=2\pi\cdot1.6~\mbox{THz}$, is in good agreement with $\Omega_0=2\pi\cdot1.7~\mbox{THz}$ of section \ref{sec:tors}. For BPy3 the agreement between fitted and calculated curve is worse, but still acceptable. The same tendency holds for the frequencies as the fitted frequency $\Omega_0^\prime=2\pi\cdot1.4~\mbox{THz}$ recovers only the right order of magnitude of $\Omega_0=2\pi\cdot2.3~\mbox{THz}$.

\subsection{Current change caused by THz radiation}

Combining the equations of section \ref{sec:THz} with the results of the present section one obtains the following effects induced by an external THz radiation for our biphenyl-like junctions if one uses the $\cos^2(\phi)$ law.

The maximal amplitudes at resonance are $\Phi_{\mbox{\tiny{max}}}^{\mbox{\tiny{BPy3}}} = 25.9^\circ$ and $\Phi_{\mbox{\tiny{max}}}^{\mbox{\tiny{BPy2}}} = 24.0^\circ$. For BPh one cannot see any effect at all as its is hardly influenced by external fields (see section \ref{sec:Influence of external electric fields}).  Whereas the amplitude of BPy2 is certainly within the range where a harmonic approximation of the energy potential surface is allowed, this is already questionable for BPy3 and anharmonic effects might become important (see also section \ref{subsec:Ground state geometry and energy}).

\begin{figure}[b]
  \centering
  \includegraphics[width=0.5\textwidth, keepaspectratio]{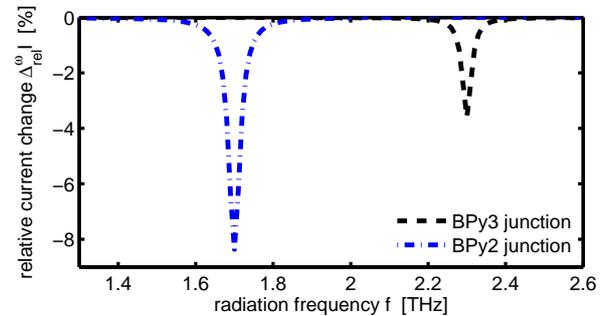}
  \caption{Current change through BPy3 and BPy2 junction depending on external radiation frequency $f=\frac{\omega}{2\pi}$.}
  \label{result}
 \end{figure}

The corresponding current changes at resonance are $\Delta_{\mbox{\tiny{rel}}}^{\omega=\Omega_0}I^{\mbox{\tiny{BPy3}}} = -3.6\%$ and $\Delta_{\mbox{\tiny{rel}}}^{\omega=\Omega_0}I^{\mbox{\tiny{BPy2}}} = -8.4\%$.  Figure \ref{result} shows for both molecules the relative change for frequencies around the resonant frequency.  For the best-fit dependence of conductance on the torsional angle (i.e. $G\propto\left|\cos(\phi)\right|^{\alpha} $, see section \ref{sec:Cos2law}) the values would be $\Delta_{\mbox{\tiny{rel}}}^{\omega=\Omega_0}I^{\mbox{\tiny{BPy3}}} = -0.6\%$ and $\Delta_{\mbox{\tiny{rel}}}^{\omega=\Omega_0}I^{\mbox{\tiny{BPy2}}} = -10.3\%$.

\subsection{Influence of geometry on sensitivity of conductance with respect to THz radiation}
\label{sec:Optimal angle}

The considerable difference in the strength of the effect on conductance between BPy3 and BPy2 can mainly traced back to their optimized torsional angle: The absolute current change at resonance,
\begin{equation}
\Delta_{\mbox{\tiny{abs}}}^{\omega =\Omega_0}I 
\propto \frac{\cos (2\phi_0)}{2} \left( \mbox{\large \it J}_0 (2\Phi_{\mbox{\tiny{max}}})-1\right),
\label{chapter4absolutchange}
\end{equation} 
is governed by the resonance amplitude $\Phi_{\mbox{\tiny{max}}}$, which is determined by eq. (\ref{Phi max}) as well as the optimized torsional angle $\phi_0$. For a given amplitude the absolute change is largest for planar (i.e. $\phi_0=0^\circ$ or $\phi_0=180^\circ$) and perpendicular ($\phi_0=90^\circ$) molecules, whereas $\phi_0$ values of $45^\circ$ or $135^\circ$ will result in zero effect.

The relative current change defined in eq. (\ref{basic relative current change equation}) will consequently be maximized for perpendicular molecules as there the zero-field current is zero or -- as the offset of the $\cos^2(\phi)$ law cannot be neglected anymore -- at least minimal.  In summary the ``perfect" molecule for a THz-sensitive molecular junction joins a perpendicular structure, low offset current combined with high total dipole moment and low damping of a torsional oscillation.
 
\begin{figure}[htb]
  \centering
  \includegraphics[width=0.25\textwidth, keepaspectratio]{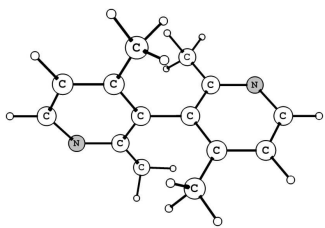}
  \caption{2,2',4,4'-tetramethyl-3,3'-bipyridine (BPyTM)}
  \label{BPyTM}
\end{figure}

Mishchenko et.\ al.\ used four methyl groups next to the inter-ring bond to get a biphenyl-like molecules with almost perpendicular geometry \cite{doi:10.1021/nl903084b}. This molecule cannot be applied immediately due to the same issues as BPh. But we can again exchange carbon atoms of the rings by nitrogen and obtain 2,2',4,4'-tetramethyl-3,3'-bipyridine (BPyTM) as shown in fig. \ref{BPyTM}. Our DFT calculations show indeed a almost perfectly perpendicular torsional angle $\phi_0=90.2^\circ$. By applying the method described in this paper one obtains $\Delta_{\mbox{\tiny{rel}}}^{\omega =\Omega_0}I\approx80\%$ , i.e. the current almost doubles due to the radiation.  However, and this should be emphasized, this value is much less reliable as for BPy3 and BPy2, because BPyTM does not allow for a clear rotation of the pyridine rings without major deformation due to the additional methyl groups. Nevertheless it demonstrates the advantage of biphenyl-like molecules with perpendicular rings.

\section{Conclusions}
\label{sec:conclusions}

{
 A model to theoretically investigate THz field induced torsional vibrations in biphenyl-like molecular junctions and their influence on the electrical transport properties has been proposed and analyzed. 
 The model was applied to biphenyl, 2,2'-bipyridine, 3,3'-bipyridine and 2,2',4,4'-tetramethyl-3,3'-bipyridine. 
 It was shown that, although for biphenyl itself the external electric field hardly couples to the torsional vibrations, for other considered systems it leads to a resonant increase of its angular vibrational amplitude by up to $25 ^\circ$. 
 The torsional vibration in turn affects the conductance through this junction according to a $\cos^2(\phi)$ law, leading to changes in conductance of order 10\% in 2,2'-bipyridine. 
 This effect can be effectively enhanced by substituting bipyridine with other junction molecules satisfying the conditions of large dipole moment in the ring plane, perpendicular to the bond connecting the two rings, and close to $90^\circ$ equilibrium torsional angle. 
 We expect such systems and the corresponding effects to be an interesting and promising direction of theoretical and experimental research in the context of molecular electronics.
 }

This work was funded by Deutsche Forschungsgemeinschaft within the Priority Program SPP 1243 and Research Training School GRK 1570. D.A.R. acknowledges support from the German Excellence Initiative via the Cluster of Excellence 1056 “Center for Advancing Electronics Dresden” (cfAED).

\providecommand{\WileyBibTextsc}{}
\let\textsc\WileyBibTextsc
\providecommand{\othercit}{}
\providecommand{\jr}[1]{#1}
\providecommand{\etal}{~et~al.}


\end{document}